\documentclass[12pt]{article}
\usepackage{amsmath,axodraw,cite,epsf}

\begin{document}

\begin{titlepage}
\begin{flushright}
TTP99-11\\
February 1999\\
\end{flushright}
\vskip 1cm
\begin{center}
{\Large  \bf 
The second order QCD contribution to
the semileptonic $b \rightarrow u$ decay rate
 } \\[20mm]
       Timo van Ritbergen \\ [2mm]
\vskip 0.5cm {\it  Institut f\"ur Theoretische Teilchenphysik,
                  Universit\"at Karlsruhe\\ D-76128 Karlsruhe, GERMANY }\\
\end{center}
\vskip 3cm
\hrule
\begin{abstract}
The order $\alpha_s^2$ contribution to the inclusive
  semileptonic decay width of a $b$ quark 
    $ \Gamma( b \rightarrow X_u e \overline{\nu}_e )$ is calculated
       analytically for zero mass $u$ quarks
\end{abstract}
\hrule
\vspace{2mm}

\end{titlepage}

\setcounter{footnote}{0} \setcounter{page}{2} \setcounter{section}{0}
\newpage


 Measurements of the semileptonic B meson decay rates
 at B factories \cite{bfactories} provide precise means to extract the
 Cabibbo-Kobayashi-Maskawa matrix elements $|V_{ub}|$ and $|V_{cb}|$
 from experiment. The accurate determination of these parameters is a 
 requisite for many tests of CP violation within the three-generation 
 Standard Model.

 The Heavy Quark Expansion provides a framework for the systematic 
 calculation of contributions to the inclusive B meson decay rate
 \cite{HQE1,HQE2,powercorrections1,powercorrections2,powercorrections3}.
 To leading order in the heavy quark mass the inclusive B meson decay rate 
  is equal to the decay rate of a on-shell $b$ quark treated within 
  renormalization group improved perturbative QCD.
  The non-perturbative corrections are suppressed by at least two powers 
  of the heavy quark mass and can be expressed as matrix elements of higher 
  dimension operators in the heavy quark effective field theory.

Perturbative QCD contributions to $b$ quark decay play an 
important role in accurate predictions for the inclusive B meson decay rate.
The first order QCD corrections to the
inclusive semileptonic $ b \rightarrow u $ decay width
   $ \Gamma( b \rightarrow X_u e \overline{\nu}_e )$  were derived  
from the calculation of the 1-loop QED corrections 
to the muon decay rate \cite{oneloopmuon}.
For $ \Gamma( b \rightarrow X_c e \overline{\nu}_e) $ 
 $c$ quark mass corrections are important and the 1-loop QCD corrections
    were obtained in a closed analytical form in Ref. \cite{nir1}.

For dimensional reasons $ \Gamma( b \rightarrow X_u e \overline{\nu}_e )$
is proportional to the $b$ quark mass to the fifth power. 
The adopted definition for the $b$ quark mass therefore 
strongly affects the coefficients of the higher order QCD corrections.
If the width is expressed in terms of the $\overline{\rm MS}$ renormalized 
   $b$ quark mass one finds that the 1-loop QCD corrections to 
         $ \Gamma( b \rightarrow X_u e \overline{\nu}_e )$  are
 significant, one-third of the size of the tree level contribution. 

  The 2-loop QCD contribution to 
   $ \Gamma( b \rightarrow X_u e \overline{\nu}_e )$  and
    $ \Gamma( b \rightarrow X_c e \overline{\nu}_e )$ 
      proportional to the number of light quark flavours was calculated in Refs.
   \cite{2loopnf1,2loopnf2}.
 This result was used to obtain an estimate for the 
  full 2-loop corrections for both semileptonic $b \rightarrow c$ and
    $ b \rightarrow  u $ decay \cite{2loopnf2}.
 Recently an accurate estimate of the full 2-loop correction to 
 the inclusive semileptonic $b \rightarrow c$ decay rate was obtained 
 \cite{estiBc}
 based on calculations at 3 distinct values for        
 the invariant mass of the lepton pair \cite{estiBc,Bcmin,Bcmax}.

 In this letter the 2-loop QCD corrections to the inclusive semileptonic 
  $b \rightarrow u$ decay rate are evaluated analytically.
  The starting point of the calculation is the Lagrangian

  \begin{equation}
   {\cal L}= {\cal L}_W +{\cal L}_{{\rm QCD}} 
  \end{equation}
   Here ${\cal L}_W$ is the effective weak interaction term 
   \begin{equation} \label{weakinteaction}
      {\cal L}_W= -\frac{G_F}{\sqrt{2}} V_{ub}
    \big[\bar\psi_{u}\gamma_\lambda (1-\gamma_5) \psi_b\big].
    \big[\bar\psi_{\nu_e}\gamma_\lambda (1-\gamma_5)\psi_{e}\big]
   \end{equation}
   where $\psi_u$, $\psi_b$, $\psi_{e}$ and $\psi_{\nu_e}$  are the
    wave functions for the $u$ quark, $b$ quark, electron and its
    neutrino respectively.
   ${\cal L}_{{\rm QCD}}$ is the standard QCD Lagrangian 
    for the strong interactions.
   The calculation is performed in effective 5 flavour QCD
   since the $t$ quark decouples for the process under consideration.
   The QCD correction to the semileptonic decay rate are finite
   since they are separately finite for a vector quark current 
   and the flavour non-singlet axial-vector quark current. 

   Throughout this article we use dimensional regularization 
     \cite{dimreg1,dimreg2} 
   and the standard modification of the minimal subtraction scheme 
      \cite{msscheme}, the $\overline{\rm MS}$-scheme \cite{msbar} 
    for coupling constant renormalization.
   Calculations with massive external fermions  
   are most easily  performed in the on-shell renormalization scheme 
   for fermion masses.
   Nevertheless, the final results in this letter will be expressed in terms of 
   $\overline{\rm MS}$ renormalized quark masses.
   For the treatment of the $\gamma_5$ matrix in dimensional regularization
    the technique described in Ref. \cite{laringamma5} 
  is used which is based on the original 
    definition of $\gamma_5$  in Ref. \cite{dimreg2}. 
   It is interesting to note that in the limit of massless 
    $u$ quarks the contribution to the decay rate coming from 
    the axial-vector part of the weak interaction Eq. (\ref{weakinteaction}) 
    can be reduced to the vector part by using
   the effective anticommutation property of the  $\gamma_5$ matrix. 
 
    For the order $\alpha_s^2$ corrections to 
  the inclusive width $\Gamma( b \rightarrow X_u e \overline{\nu}_e )$
  one has to determine the sum of the decay widths   
     $  b \rightarrow u e \overline{\nu}_e $,
        $  b \rightarrow u e \overline{\nu}_e g $,
        $  b \rightarrow u e \overline{\nu}_e g g$ and
         $  b \rightarrow u e \overline{\nu}_e q \overline{q}$, where 
            $g$ denotes a gluon and $q$ denotes a quark,
         with up to two gluons or quarks in virtual corrections.
  According to the Kinoshita-Lee-Nauenberg \cite{kln} theorem, the inclusive
   decay rate is free from singularities as the $u$ quark mass goes
   to zero. The calculation will be done in the approximation
   of massless $u$,$d$,$s$ and $c$ quarks.
   The $c$ quark enters only in the partial rate
       $  b \rightarrow u e \overline{\nu}_e c \overline{c}$
   or as $c\overline{c}$ pairs in virtual corrections, 
   and the effect of a non-zero $c$ quark mass on the total decay rate 
      is therefore limited.
   The situation is quite different for
    $\Gamma( b \rightarrow X_c e \overline{\nu}_e )$ where the $c$ quark
   enters in all diagrams and $c$ quark mass effects are very important.

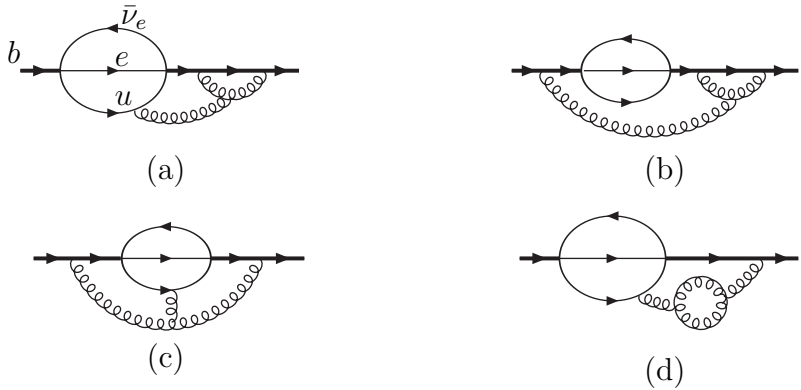
\begin{figure}
\begin{center}
\hfill
\begin{picture}(120,70)(0,0)
 \SetWidth{1.5}
 \Line(15,20)(30,20)
 \Line(70,20)(120,20)
 \SetWidth{0.8}
 \ArrowLine(15,20)(30,20)
 \ArrowLine(70,20)(84,20)
 \ArrowLine(84,20)(107,20)
 \ArrowLine(107,20)(120,20)
 \SetWidth{0.5}
 \ArrowLine(30,20)(70,20)
 \Oval(50,20)(16,20)(0)
 \ArrowLine(50.01,36)(49.01,36)
 \ArrowLine(49.99,4)(50.01,4)
 \GlueArc(94.8,23)(12,195,345){2.0}{7.5}
 \GlueArc(72,36)(34,246,310){2.0}{9.5}
 \Text(10,23)[bl]{$b$}
 \Text(50.8,22)[bl]{$e$}
 \Text(53,36)[bl]{$\bar\nu_e$}
 \Text(50.8,7)[bl]{$u$}
     \Text(70,-12)[t]{(a)}
\end{picture}
\hfill
\begin{picture}(120,70)(0,0)
 \SetWidth{1.5}
 \Line(12,20)(38,20)
 \Line(72,20)(120,20)
 \SetWidth{0.8}
 \ArrowLine(12,20)(25,20)
 \ArrowLine(28,20)(38,20)
 \ArrowLine(72,20)(84,20)
 \ArrowLine(84,20)(107,20)
 \ArrowLine(107,20)(120,20)
 \SetWidth{0.5}
 \ArrowLine(39.2,20)(72.8,20)
 \Oval(55,20)(12,16.8)(0)
 \ArrowLine(56.01,32)(54.01,32)
 \ArrowLine(54.99,8)(56.01,8)
 \GlueArc(94.8,23)(12,195,345){2.0}{7.5}
 \GlueArc(64,45)(48,212,312){2.0}{18.5}
 \Text(70,-12)[t]{(b)}
\end{picture}
\hfill\null\\
\hfill
\begin{picture}(120,70)(0,0)
 \SetWidth{1.5}
 \Line(20,20)(52.8,20)
 \Line(86.8,20)(122,20)
 \SetWidth{0.8}
 \ArrowLine(20,20)(35,20)
 \ArrowLine(35,20)(52.8,20)
 \ArrowLine(86.8,20)(106,20)
 \ArrowLine(106,20)(122,20)
 \SetWidth{0.5}
 \ArrowLine(53.2,20)(86.8,20)
 \Oval(70,20)(12,16.8)(0)
 \ArrowLine(70.01,32)(69.01,32)
 \ArrowLine(69.99,8)(70.01,8)
 \GlueArc(70,32)(37,199,341){2.0}{20.5}
 \Gluon(72,8)(72,-4){2.0}{2.3}
 \Text(70,-12)[t]{(c)}
\end{picture}
\hfill
\begin{picture}(120,70)(0,0)
 \SetWidth{1.5}
 \Line(15,20)(30,20)
 \Line(70,20)(120,20)
 \SetWidth{0.8}
 \ArrowLine(15,20)(30,20)
 \ArrowLine(70,20)(108,20)
 \ArrowLine(108,20)(120,20)
 \SetWidth{0.5}
 \GlueArc(83.2,3.2)(8,0,385){2.0}{11.0}
 \ArrowLine(30,20)(70,20)
 \Oval(50,20)(16,20)(0)
 \ArrowLine(49.99,4)(50.01,4)
 \ArrowLine(50.01,36)(49.99,36)
 \GlueArc(76,36)(34,242,266){2.0}{3.5}
 \GlueArc(76,36)(34,298,331){2.0}{4.5}
 \Text(70,-17)[t]{(d)}
\end{picture}
\hfill\null\\
\vglue 18pt
\end{center}
\caption{Examples of non-abelian diagrams whose cuts give contributions to
 $  b \rightarrow u e \overline{\nu}_e $,
        $  b \rightarrow u e \overline{\nu}_e g $,
        $  b \rightarrow u e \overline{\nu}_e g g$. }
\label{fig1}
\end{figure}

   Using the optical theorem one may express the inclusive
   decay rate as the imaginary part of 4-loop on-shell
   propagator type diagrams. Some of these diagrams are shown in Fig 1.
   The imaginary part of these 4-loop quark propagator diagrams 
   can be evaluated analytically using the methods employed in 
     Ref. \cite{muon2loop}
   where the 2-loop QED corrections to the muon decay rate were
   evaluated. As was noted before, the muon decay rate and 
    $\Gamma( b \rightarrow X_u e \overline{\nu}_e )$ are closely
   related; the result for the 2-loop QED corrections to
     the muon decay width corresponds 
   formally to keeping only the abelian part in the 2-loop QCD corrections 
   to $\Gamma( b \rightarrow X_u e \overline{\nu}_e )$.
   The approach of Ref. \cite{muon2loop} relies heavily on the 
     use of the method of 
   integration-by-parts \cite{ibp} within dimensional regularization
   to express the imaginary part of the 4-loop diagrams in terms
   of a small set of 26 primitive integrals. Although there are
   diagrams for $\Gamma( b \rightarrow X_u e \overline{\nu}_e )$
   that involve integration topologies that were absent in 
    Ref. \cite{muon2loop} 
   it turns out that after the use of integration-by-parts identities  
   for these topologies no new primitive integrals are needed.

   All diagrams that contribute to 
   $\Gamma( b \rightarrow X_u e \overline{\nu}_e )$
   are calculated in a general covariant gauge for the 
   gluon fields. The explicit cancellation of the gauge dependence in
   the sum of the diagrams gives an important check of the result. 
   The result that is obtained in this way reads

 \begin{equation} \label{mainrespolebasic}
       \Gamma( b \rightarrow X_u e \overline{\nu}_e ) =
        \frac{ G_F^2  |V_{ub}|^2 M_b^5 }{192 \pi^3} \left[
        1 + b_1  \frac{ \alpha_s^{(5)}(\mu) }{\pi}
          + b_2 \left(  \frac{\alpha_s^{(5)}(\mu) }{\pi} \right)^2 
      + O( \alpha_s^3 )      \right]
          \hspace{1cm}  \end{equation}
 \begin{eqnarray}
 b_1 & = & \textstyle C_F \left( \frac{25}{8} - 3 \zeta_2 \right)  
       \nonumber \\
      & \approx & - 2.41307   \nonumber \\
 b_2 & = & \textstyle
        C_A C_F  \left[
           \frac{154927}{10368}
          - \frac{53}{2} \zeta_2 \ln(2)
          + \frac{95}{27} \zeta_2
          - \frac{383}{72} \zeta_3
          + \frac{101}{16} \zeta_4
          + \frac{275}{96} \ln\left(\frac{ \mu^2 }{M_b^2} \right)
          - \frac{11}{4} \zeta_2  \ln\left(\frac{ \mu^2 }{M_b^2} \right)
          \right]  \nonumber \\
  &  & \textstyle
       + C_F^2  \left[
           \frac{11047}{2592}
          + 53 \zeta_2 \ln(2)
          - \frac{1030}{27} \zeta_2
          - \frac{223}{36} \zeta_3
          + \frac{67}{8} \zeta_4
          \right]   \nonumber \\
   &    &  \textstyle
       + C_F T_F n_f  \left[
          - \frac{1009}{288}
          + \frac{77}{36} \zeta_2
          + \frac{8}{3} \zeta_3
          - \frac{25}{24} \ln\left(\frac{ \mu^2 }{M_b^2} \right)
          + \zeta_2 \ln\left(\frac{ \mu^2 }{M_b^2} \right) 
          \right]   \nonumber \\
  &  & \textstyle + C_F T_F \left(
           \frac{6335}{192}
          - \frac{9}{2} \zeta_2
          - 24 \zeta_3
          \right)  \nonumber \\ 
  & \approx &\textstyle
         - 21.29553 
         - 4.625050 \ln\left(\frac{ \mu^2 }{M_b^2} \right)
 \label{mainrespole}   \end{eqnarray}
 where $\alpha_s^{(5)}(\mu)$ is the coupling constant in effective
 5 flavour QCD, $\mu$ is the renormalization scale,
    $n_f=5$ is the total number of quark flavours and
  $M_b$ is the  $b$ quark pole mass in the two loop order.
 $C_F = 4/3$  and $C_A = 3$ are the Casimir operators of the 
 fundamental and adjoint representation of the colour group SU(3),
 $T_F=1/2$ is the trace normalization of the fundamental representation.
 $\zeta_k$ denotes the Riemann zeta function, 
    $\zeta_2 =  \pi^2/6 $,
    $\zeta_3 = 1.2020569\cdots $,
    $\zeta_4 =  \pi^4/90$.

   We note that  the coefficient of $C_F T_F n_f$
     in  Eq.(\ref{mainrespole})  
   agrees with the value $3.22$ obtained in Ref. \cite{2loopnf1} 
   and the terms proportional to $C_F^2$    
   and $C_F T_F$ agree with the result for the QED contributions
     to the muon decay width of Ref. \cite{muon2loop}.
   The order $\alpha_s^2$ coefficient
   in Eq. (\ref{mainrespole}) is sizable but this coefficient 
   is somewhat smaller than the estimate $ -28.7 $ obtained in Ref. 
    \cite{2loopnf2}.

   It has been shown \cite{renormalon1,renormalon2} that 
    the perturbative coefficients for
   $\Gamma( b \rightarrow X_u e \overline{\nu}_e ) $ 
   receive contributions that grow rapidly at higher orders
   corresponding to a singularity  in the Borel plane at 
   $\pi/(2\beta_0)$ (with $\beta_0$ given below Eq. (\ref{beta}))
   when the decay width is expressed in terms of the $b$ quark pole mass. 
   This particular growth of the perturbative coefficients is related to 
    the infrared sensitivity of the pole mass definition and is absent 
   when the inclusive decay width is expressed in terms of
   the $\overline{\rm MS}$ renormalized $b$ quark mass $m_b(\mu)$.
   We will therefore express the decay width in terms of $m_b(\mu)$ 
   using the known relation between the pole quark mass and the
    $\overline{\rm MS}$ renormalized quark mass \cite{polemsbar1,polemsbar2}.
 \begin{equation} \label{mbarpolemain}
    M_b = m_b(\mu) \left[ 1 + c_1  \frac{ \alpha_s^{(5)}(\mu) }{\pi}
             + c_2 \left( \frac{ \alpha_s^{(5)}(\mu) }{\pi} \right)^2
             + O( \alpha_s^3 ) \right]  \hspace{4cm}  \end{equation}
\begin{eqnarray}
   c_1 & = & \textstyle
      C_F \left( 1 + \frac{3}{4} \ln\left(\frac{ \mu^2 }{M_b^2} \right) 
            \right)  \nonumber \\
  c_2  & = &  \textstyle
      C_A C_F \left[
           \frac{1111}{384}
          + \frac{3}{2} \zeta_2 \ln(2)
          - \frac{1}{2} \zeta_2
          - \frac{3}{8} \zeta_3
          + \frac{185}{96}  \ln\left(\frac{ \mu^2 }{M_b^2} \right) 
          + \frac{11}{32}  \ln^2\left(\frac{ \mu^2 }{M_b^2} \right) 
          \right]  \nonumber \\
  & & \textstyle
       + C_F^2  \left[
           \frac{121}{128}
          - 3 \zeta_2 \ln(2)
          + \frac{15}{8} \zeta_2
          + \frac{3}{4} \zeta_3
          + \frac{27}{32}  \ln\left(\frac{ \mu^2 }{M_b^2} \right) 
          + \frac{9}{32} \ln^2\left(\frac{ \mu^2 }{M_b^2} \right)
          \right]  \nonumber  \hspace{3cm} \\
  & & \textstyle
      + C_F T_F n_f \left[
          - \frac{71}{96}
          - \frac{1}{2} \zeta_2
          - \frac{13}{24}  \ln\left(\frac{ \mu^2 }{M_b^2} \right)
          - \frac{1}{8} \ln^2\left(\frac{ \mu^2 }{M_b^2} \right)
          \right]
       + C_F T_F \left(
          - \frac{3}{4}
          + \frac{3}{2} \zeta_2
          \right)   \label{mbarpole}
\end{eqnarray}
Applying this relation to eliminate $M_b$ from Eq.(\ref{mainrespolebasic}) 
and from the logarithms appearing in Eqs. (\ref{mainrespole},\ref{mbarpole}) 
\begin{equation}  
         \ln\left(\frac{ \mu^2 }{M_b^2} \right)
        =  \ln\left(\frac{ \mu^2 }{m_b^2(\mu)} \right) 
       -2 C_F \left[ 1 +\frac{3}{4} \ln\left(\frac{ \mu^2 }{m_b^2(\mu)} 
            \right) \right]  \frac{ \alpha_s^{(5)}(\mu) }{\pi}
          + O( \alpha_s^2 )  \hspace{1cm}
 \end{equation}
one obtains the result for the decay width in terms of 
 the  $\overline{\rm MS}$ renormalized $b$ quark mass
 \begin{equation} \label{mainresmbar5basic}
      \Gamma( b \rightarrow X_u e \overline{\nu}_e ) =
    \frac{ G_F^2  |V_{ub}|^2  }{192 \pi^3} 
     \left( m_b^{(5)}(\mu)\right)^5 \left[
        1 + h_1  \frac{ \alpha_s^{(5)}(\mu) }{\pi}
          + h_2 \left(  \frac{\alpha_s^{(5)}(\mu) }{\pi} \right)^2
      + O( \alpha_s^3 )      \right]
 \end{equation}
 \begin{eqnarray}
  h_1 & = & \textstyle
        C_F \left[
           \frac{65}{8}
          - 3 \zeta_2
          + \frac{15}{4} \ln\left(\frac{ \mu^2 }{m_b^2(\mu)} \right) 
          \right]  \nonumber \\
  &  \approx & \textstyle
           4.25360 + 5 \ln\left(\frac{ \mu^2 }{m_b^2(\mu)} \right)
                   \nonumber \\
  h_2  & =  & \textstyle
       C_A C_F \left[
           \frac{19057}{648}
          - 19 \zeta_2 \ln(2)
          + \frac{55}{54} \zeta_2
          - \frac{259}{36} \zeta_3
          + \frac{101}{16} \zeta_4
          + \frac{25}{2}  \ln\left(\frac{ \mu^2 }{m_b^2(\mu)} \right) 
          \right.  \nonumber \\
    & & \textstyle \left. \;\;\;\;
          - \frac{11}{4} \zeta_2  \ln\left(\frac{ \mu^2 }{m_b^2(\mu)} \right)
          + \frac{55}{32}  \ln^2\left(\frac{ \mu^2 }{m_b^2(\mu)} \right)
          \right]  \nonumber \\
    & &  \textstyle
      + C_F^2 \left[
           \frac{281113}{10368}
          + 38 \zeta_2 \ln(2)
          - \frac{9455}{216} \zeta_2
          - \frac{22}{9} \zeta_3
          + \frac{67}{8} \zeta_4
          + \frac{405}{16}  \ln\left(\frac{ \mu^2 }{m_b^2(\mu)} \right)
    \right.  \nonumber \\
  & & \textstyle \left. \;\;\;\;
          - \frac{45}{4} \zeta_2  \ln\left(\frac{ \mu^2 }{m_b^2(\mu)} \right)
          + \frac{225}{32}  \ln^2\left(\frac{ \mu^2 }{m_b^2(\mu)} \right)
          \right]  \nonumber \\
   & & \textstyle
          + C_F T_F n_f \left[
          - \frac{1037}{144}
          - \frac{13}{36} \zeta_2
          + \frac{8}{3} \zeta_3
          - \frac{15}{4} \ln\left(\frac{ \mu^2 }{m_b^2(\mu)} \right)
          + \zeta_2 \ln\left(\frac{ \mu^2 }{m_b^2(\mu)} \right)
          - \frac{5}{8} \ln^2\left(\frac{ \mu^2 }{m_b^2(\mu)} \right) 
          \right]  \nonumber \\
      & &  \textstyle
      + C_F T_F \left(
           \frac{5615}{192}
          + 3 \zeta_2
          - 24 \zeta_3
          \right)  \nonumber \\
   &  \approx & \textstyle  
           26.78476 
         + 36.99016 \ln\left(\frac{ \mu^2 }{m_b^2(\mu)} \right)
         + 17.29167  \ln^2\left(\frac{ \mu^2 }{m_b^2(\mu)} \right) 
 \end{eqnarray}
where  $m_b(\mu) = m_b^{(5)}(\mu)$ is the $\overline{\rm MS}$ renormalized
 $b$ quark mass in effective 5 flavour QCD.
The decay width may also be expressed in terms of parameters of
    4 flavour QCD by using the known decoupling relations for 
 the coupling constant and the quark masses 
      \cite{decouple1,decouple2,decouple3}
 \begin{eqnarray}
   \frac{ \alpha_s^{(5)}(\mu) }{\pi} & = &
         \frac{ \alpha_s^{(4)}(\mu) }{\pi}
         + \left( \frac{ \alpha_s^{(4)}(\mu) }{\pi} \right)^2
        \frac{ T_F}{3} \ln\left( \frac{\mu^2}{m_b^2(\mu)}\right)
            + O (\alpha_s^3)  \\
    m_q^{(5)}(\mu)  & = &  m_q^{(4)}(\mu) \left[
        1 + \left( \frac{ \alpha_s^{(4)}(\mu) }{\pi} \right)^2
      T_F C_F \left( -\frac{1}{8} \ln^2\left( \frac{\mu^2}{m_b^2(\mu)}\right)
   \right. \right. \nonumber  \\
  & &  \left. \left.
              + \frac{5}{24}\ln\left( \frac{\mu^2}{m_b^2(\mu)}\right)
            - \frac{89}{288} \right)  + O(\alpha_s^3) \right] 
 \end{eqnarray}
to obtain
 \begin{equation}   \label{mainresmbar4basic}
      \Gamma( b \rightarrow X_u e \overline{\nu}_e ) =
    \frac{ G_F^2  |V_{ub}|^2  }{192 \pi^3}
     \left( m_b^{(4)}(\mu)\right)^5 \left[
        1 + f_1  \frac{ \alpha_s^{(4)}(\mu) }{\pi}
          + f_2 \left(  \frac{\alpha_s^{(4)}(\mu) }{\pi} \right)^2
      + O( \alpha_s^3 )      \right]
 \end{equation}
\begin{eqnarray}
     f_1 &  \approx & \textstyle
           4.25360 + 5 \ln\left(\frac{ \mu^2 }{m_b^2(\mu)} \right)
                \nonumber \\
     f_2 & \approx & \textstyle
         25.75467
        + 38.39353 \ln\left(\frac{ \mu^2 }{m_b^2(\mu)} \right)
        + 17.70833 \ln^2\left(\frac{ \mu^2 }{m_b^2(\mu)} \right)  .
 \end{eqnarray}
 Here $m_b(\mu) = m_b^{(4)}(\mu)$ and only the numerical
 expressions are given.

 Each of the obtained expressions for 
   the decay width Eqs. (\ref{mainrespolebasic}),(\ref{mainresmbar5basic})
         and (\ref{mainresmbar4basic}) 
   satisfies formal renormalization group invariance  
   that is required of a physical quantity 
    which in the $\alpha_s^2$ approximation reads 
   \begin{equation} \label{renormindep}
      \frac{d}{d \ln(\mu^2) } 
              \Gamma(\mu^2,\alpha_s(\mu),m_b(\mu)) = O(\alpha_s^3(\mu)) 
    \end{equation}
   The $\mu$ dependence in the renormalized mass and coupling constants 
   is given by the renormalization group equations 
 \begin{eqnarray}
  \label{beta} 
  \frac{d a }{d \ln \mu^2}  & = & \beta(a) 
   =  -\beta_0 a^2 - \beta_1 a^3 -\beta_2 a^4
           -\beta_3 a^5 + O(a^6) \hspace{1cm} \\
  \label{gamma}
  \frac{d\ln m_q }{d \ln \mu^2}  & = & \gamma_m(a) = 
       -\gamma_0 a - \gamma_1 a^2
       -\gamma_2 a^3 -\gamma_3 a^4 + O(a^5) \hspace{1cm}
  \end{eqnarray}
 where $a = \alpha_s/\pi$, 
    $\beta_0 = \frac{11}{12} C_A - \frac{1}{3} T_F n_f$, 
            $\gamma_0=\frac{3}{4} C_F$,
               $\gamma_1= C_F( \frac{3}{32}C_F
                 +\frac{97}{96} C_A - \frac{5}{24} T_F n_f)$
   and $ n_f $ is the number of quark flavours in effective QCD.
  The higher order coefficients up to $\beta_3$ and $\gamma_3$
   are given in \cite{renorm1}.

  As is indicated in Eq. (\ref{renormindep}) the two loop QCD expressions for
  $ \Gamma( b \rightarrow X_u e \overline{\nu}_e ) $  have a
   remaining renormalization scale dependence that is of order $\alpha_s^3$.
   The renormalization scale dependence of the expression for the
   decay width Eq. (\ref{mainresmbar5basic}) 
     in the different orders of perturbative QCD 
   is illustrated in Fig 2.
   In Fig. 2 the leading order curve (LO) corresponds to
   the tree level expression for the decay width and to solving
   the renormalization group equation for $\alpha_s(\mu)$ and $m_b(\mu)$ in the 
   leading order, the next-to-leading order curve (NLO) corresponds to keeping
   one higher order in $\alpha_s$  both in the expression for
   the decay width and the renormalization group equations, etc.
   The renormalization group equation for $\alpha_s(\mu)$ is 
   solved numerically starting from $\alpha_s^{(5)}(m_Z) = 0.118$. 
    Similarly the scale evolution of $m_b^{(5)}(\mu)$ is done numerically
     starting from $m_b^{(5)}(m_b) = {\rm 4.3\; GeV}$.
   For comparison we have included a NNNLO curve that corresponds to
   taking the presently unknown constant of order $\alpha_s^3$ 
   in  Eq. (\ref{mainresmbar5basic}) to be $5^3 =125 $, assuming here
   an approximate power-like growth of the lower order coefficients, see e.g. 
    Ref.\cite{nequal5}.
   The logarithmic terms $ \alpha_s^3 \ln^i(\mu^2/m_b^2(\mu))$, $i=1,2,3$ 
    can be expressed in terms of the known lower order
   coefficients by requiring formal $\mu$-independence 
   of the decay rate in the $\alpha_s^3$ order.

\begin{figure}
\hspace{2.5cm}  \epsfxsize=11.5cm \epsfbox{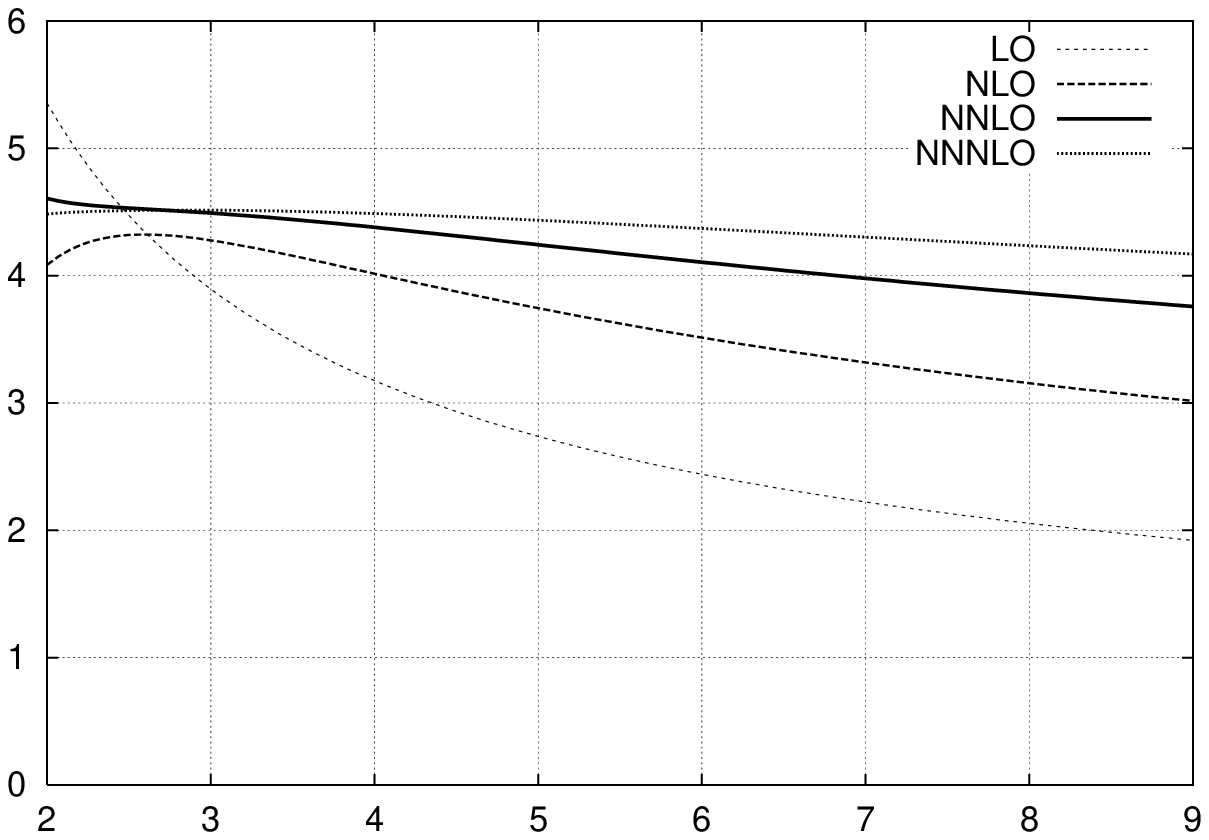}

\vspace{-5.0cm}  \hspace{2.3cm} \begin{picture}(0,0)(0,0)
           \rText(0,0)[r][l]{
           $\Gamma \; [10^{-7}\, {\rm eV}]$ }
           \end{picture}

\vspace{4.3cm}
\hspace{8cm}  $\mu \;[ {\rm GeV}]$

\caption{ Renormalization scale dependence of
       $ \Gamma( b \rightarrow X_u e \overline{\nu}_e ) $
       near $\mu = m_b(m_b)$
       in different orders of perturbative QCD
       for $\alpha_s^{(5)}(m_Z) = 0.118$,
        $m_b^{(5)}(m_b) = {\rm 4.3\; GeV}$ and $|V_{ub}| = 0.003 $ . }

\end{figure}

    One may see from Fig. 2 that in the NNLO expression 
      Eq. (\ref{mainresmbar5basic}) a moderate 
     renormalization scale dependence remains near $\mu = m_b(m_b)$.
    Variation of $\mu$ in the relatively wide interval between 
      $\mu = 1/2 \, m_b(m_b)$ and $\mu = 2\, m_b(m_b)$ changes the decay 
      width in the NNL order by about 15\%.
  If variation with respect to $\mu$ in this interval is used to estimate the
 contributions to $\Gamma(b \rightarrow X_u e \overline{\nu}_e )$
  beyond the $\alpha_s^2$ order one finds that these
    unknown higher order contributions should not
     hinder the extraction of $|V_{ub}|$ down to an accuracy of 8\%.
   Furthermore we note that the difference
     between the LO,NLO and NNLO 
     curves is minimal for $\mu \approx {\rm 2.5\; GeV}$, 
     i.e. the apparent convergence of the perturbation series is optimal
     for this choice of the renormalization scale
     (which is within the above mentioned interval).
     This fact is not surprising as the size of the 
      coefficients $h_1$ and $h_2$ in Eq. (\ref{mainresmbar5basic})
      is greatly reduced for this value of $\mu$.
     It is also in agreement with the observation \cite{nequal5}
       that a more natural renormalization scale 
     for the $\overline{\rm MS}$ renormalized  $b$ quark mass 
     in inclusive semileptonic $b$ decay is somewhat below $m_b(m_b)$
     as the scale is set by the characteristic energy that is released
     into the hadronic final state.
     Adopting $\mu \approx {\rm 2.5\; GeV}$ as a choice for the
     renormalization scale we conclude that the  
     $\alpha_s^2$ order approximation to 
      $\Gamma(b \rightarrow X_u e \overline{\nu}_e )$
      that is obtained in this
     letter provides a good theoretical foundation for the
     extraction of $|V_{ub}|$ from inclusive semileptonic B meson decays.

 Useful discussions with  K. Chetyrkin, A. Czarnecki, Th. Mannel,
  K. Melnikov and Y.-P. Yao  are gratefully acknowledged. 
 This work was supported in part by BMBF under contract No. 057KA92P and
 DFG Forschergruppe under contract KU 502/8-1.

\end{document}